\documentclass[pre,aps,twocolumn]{revtex4}
\input epsf

\usepackage{amsmath}
\usepackage{amssymb}
\usepackage{graphicx}
\usepackage{latexsym}

\begin{document}

\title{Statistical properties of sampled networks}

\author{Sang Hoon Lee}
 \email{lshlj@stat.kaist.ac.kr}
\author{Pan-Jun Kim}
 \email{extutor@gmail.com}
\author{Hawoong Jeong}
 \email{hjeong@kaist.edu}
\affiliation {Department of Physics, Korea
Advanced Institute of Science and Technology, Daejeon 305-701,
Korea}

\date{\today}

\begin{abstract}
We study the statistical properties of the sampled scale-free
networks, deeply related to the proper identification of various
real-world networks. We exploit three methods of sampling and
investigate the topological properties such as degree and
betweenness centrality distribution, average path length,
assortativity, and clustering coefficient of sampled networks
compared with those of original networks. It is found that the
quantities related to those properties in sampled networks appear to
be estimated quite differently for each sampling method. We explain
why such a biased estimation of quantities would emerge from the
sampling procedure and give appropriate criteria for each sampling
method to prevent the quantities from being overestimated or
underestimated.
\end{abstract}

\maketitle

\section{introduction}
Recently, a huge amount of research on complex networks has been
achieved in interdisciplinary fields including mathematics,
statistical physics, computer science, sociology, biology,
etc.~\cite{Newman2003a,Albert2002,Dorogovtsev2001a}. Complex
networks are ubiquitous in the real world, e.g., there are
technological networks such as the Internet~\cite{Faloutsos1999},
biological networks such as protein interaction
networks~\cite{Jeong2001}, and social networks such as scientific
collaboration networks~\cite{Newman2001a}. Various models to explain
the observed properties of those real networks have been introduced
and studied by both numerical and analytic approaches. Relatively
fewer works, however, have been done about possible error or bias in
collecting data and identifying real networks in a practical sense,
and most works deal exclusively with either social networks or the
Internet~\cite{Costenbader2003,Robins2004,Kossinets2004,Petermann2004b,
Clauset2004c,Achlioptas2005,Dallasta2005,Stumpf2005,Scholz2005,
JDHan2005,Stumpf2005c}.

For instance, a survey of relationships among participants has to be
conducted to construct a social network, but the collected
network data may be incomplete or erroneous since a survey usually
targets only a partial sample of a whole
population~\cite{Kossinets2004}. The topology of the Internet is
inferred by aggregating paths or {\em traceroutes}~\cite{Traceroute},
which also reveals only a part of the whole
Internet~\cite{Petermann2004b,Clauset2004c,Achlioptas2005,Dallasta2005}.
In biology, protein-protein interaction networks are identified by
seeking contextual or cellular functions mostly within specific functional
modules~\cite{Jeong2001}. Identification of such networks by
experiments also has a fundamental limit naturally. Thus, all
these networks identified are {\em sampled} networks from complete
structures. In addition, if the size of an entire network is too
large to measure some quantities such as betweenness
centrality~\cite{KIGoh2001,KIGoh2002} due to time complexity,
inevitably a sampling process is necessary.

So far models of networks have been designed based on features
observed in real networks, such as the small-world
effect~\cite{Watts1998} and the power-law degree
distribution~\cite{Barabasi1999,Krapivsky2001}. But what if those
{\em observed} characteristics from the {\em sampled} networks are
considerably different from the original structures of the real
networks? It has been shown that the sampled networks based on the
traceroute sampling method may have significantly different
topological properties from the original network in some
cases~\cite{Petermann2004b,Clauset2004c,
Achlioptas2005,Dallasta2005}.
Effects of missing data in social networks are discussed in
Ref.~\cite{Kossinets2004}, in which it was shown that some problems
in conceiving social networks can cause incompleteness of data and
lead to misestimation of quantities like mean node degree,
clustering coefficient, assortativity, etc. At this point, bias in
such quantities needs to be considered in a more general sense.

In a statistical sense, inference from a sample provides fairly
reasonable estimation of a whole population if a large number of
objects are selected randomly enough to be representative in the
population. This naive criterion, however, cannot be applied
directly to sampling networks, since there are two different
elements, i.e., nodes and links in a network. A degree distribution
of nodes is, for example, a statistic of a network, but the degree
is not an independent characteristic of each node. Nodes are
literally connected to one another, by the other kind of components
called links from which a degree is defined. Similarly, other
properties of a network also heavily depend on the way that nodes
and links are interwoven. There could be several different ways of
sampling networks due to the two interrelated elements (nodes and
links), and each method may give distinctive features with respect
to such properties.

There has been a large amount of work on random breakdowns or
intentional attack on complex networks, considered as the exact
reverse process of sampling, in the physics
community~\cite{Albert2000a,Cohen2000,Cohen2000a, Gallos2005b}. The
analytic methods in that work, therefore, can be also applied to the
sampling problem. In this paper, we adopt three basic methods of
sampling networks and investigate the effect of each method on
measuring several well-known network quantities such as degree
distribution, average path length, betweenness centrality
distribution~\cite{KIGoh2001,KIGoh2002},
assortativity~\cite{Newman2002a}, and clustering
coefficient~\cite{Watts1998}. Observed bias of such quantities is
explained, and we provide appropriate criteria for choosing sampling
methods to measure the quantities more accurately. Some typical real
networks as well as the Barab{\'a}si-Albert
model~\cite{Barabasi1999} are sampled for this analysis. More
general sampling processes used to identify real networks may
consist of some combinations of methods presented here or variations
of them, but we can infer by using the results from the basic
methods.

\section{sampling methods and networks}
\begin{figure}
\includegraphics[width=0.4\textwidth]{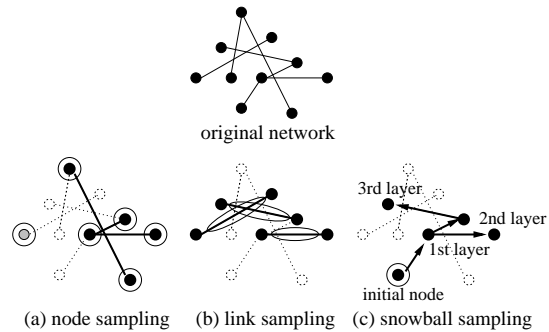}
\caption{Three kinds of sampling method. (a) Node sampling: Select
the circled nodes, keep three links among them, and the isolated
node is removed. (b) Link sampling: Select the three circled links
and six nodes attached to them. (c) Snowball sampling: Starting from
the circled node, select nodes and links attached to them by tracing
links.} \label{method}
\end{figure}

We introduce three kinds of sampling method called node sampling,
link sampling, and snowball sampling. In node sampling, a certain
number of nodes are randomly chosen and links among them are kept.
The sampling fraction in this method is defined as the ratio of the
number of chosen nodes (including isolated nodes that will be
removed later) to that of all the nodes in the original network. As
in Fig.~\ref{method}(a), isolated nodes are neglected for
convenience, although they are fully predictable, so the number of
nodes in a sampled network is a little bit less than that of
selected nodes. We observe the dependence of the number of chosen
links on that of nodes, since it is related to the average degrees
and average path length of sampled networks, discussed later on.
Suppose the fraction of number of selected nodes is $\alpha$ and
that of links among them is $\beta$. Then it is found that $\beta
\sim \alpha^2$ if we pick nodes randomly, since the maximum number
of (undirected) links possible for $n$ selected nodes are ${n
\choose 2} = n(n-1)/2 \sim n^2$~\cite{pick_node}.

In link sampling, a certain number of links are randomly selected
and nodes attached to them are kept, as in Fig.~\ref{method}(b).
In snowball sampling~\cite{snowball,Newman2003c}, we first choose a
single node and all the nodes directly linked to it are picked. Then
all the nodes connected to those picked in the last step are
selected, and this process is continued until the desired number of
nodes are sampled. The set of nodes selected in the $n$th step is
denoted as the $n$th layer, in the same sense of ``radius'' for
ego-centered networks in Ref.~\cite{Newman2003c}. See
Fig.~\ref{method} (c) for illustration. To control the number of
nodes in the sampled network, a necessary number of nodes are
randomly chosen from the last layer. Similar to the cluster-growing
method used to calculate the fractal dimension of percolation
clusters in Ref.~\cite{Song2005}, the snowball sampling method tends
to pick hubs (nodes with many links) in short step due to high
connectivity of them. So whether the initial node is a hub or not
does not make a noticeable difference in characterizing the sampled
network.

For numerical analysis of the sampled networks, we use
Barab{\'a}si-Albert (BA) scale-free network as an example of model
networks, which follows the power-law degree distribution $p(k) \sim
k^{-3}$, with $30000$ nodes and $m_0 = m = 4$~\cite{Barabasi1999}.
We also consider three real-world networks from various fields,
including protein interaction network
(PIN)~\cite{Jeong2001,KIGoh2003}, the Internet at the autonomous
systems (AS) level~\cite{Meyer}, and e-print archive coauthorship
network (arxiv.org)~\cite{Newman2001a}. The numbers of nodes and
links for each network are in Table~\ref{net_table}. Although
results from other homogeneous networks are also discussed in Sec.
IV, most of networks considered in this work are undirected and
scale-free networks following power-law degree distribution, $p(k)
\sim k^{-\gamma}$, where $2 < \gamma \leq 3$.

\begin{table}
\begin{tabular}{cccc}
\hline \hline
Network & $n$ & $l$ & Ref. \\
\hline
PIN & 5077 & 16449 & \cite{Jeong2001,KIGoh2003} \\
Internet AS & 10515 & 21455 & \cite{Meyer} \\
arxiv.org & 49983 & 245300 & \cite{Newman2001a} \\
\hline \hline
\end{tabular}
\caption{The numbers of nodes $n$ and links $l$ for each real
network.} \label{net_table}
\end{table}

\section{characteristics of sampled networks}

\subsection{Degree distribution and average path length}
\begin{figure}
\includegraphics[width=0.45\textwidth]{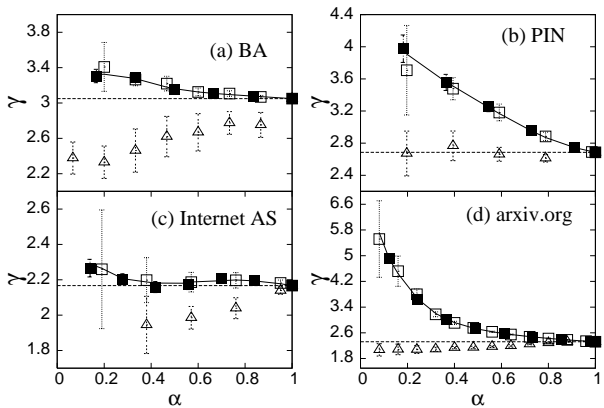}
\caption{Changes of degree exponent $\gamma$ for each network's
sampled networks according to the sampling fraction $\alpha$,
averaged over ten independent realizations. Empty squares
($\square$) stand for node sampling, filled squares ($\blacksquare$)
for link sampling, and empty triangles ($\vartriangle$) for snowball
sampling. The horizontal dashed lines are the values for the
original exponent of each network, and the solid lines represent the
values obtained by Eq.~(\ref{new_degree_node}).} \label{MLE_deg}
\end{figure}

A degree of a node is defined as the number of links attached to the
node. Many real networks are shown to have a power-law degree
distribution $p(k) \sim k^{-\gamma}$~\cite{Newman2003a,Albert2002,
Dorogovtsev2001a}, including the networks considered in this paper.
We found that in general degree distributions of sampled networks
from the four networks obtained by all three methods follow the
power-law as well as those of the original networks. The exponents
of degree distribution $\gamma$ (degree exponent) are extracted
using maximum likelihood estimate given by the
formula~\cite{Newman2004b}
\begin{equation}
\gamma = 1 + n \left[ \sum_{i=1}^n \ln \frac{k_i}{k_\textrm{min}}
\right]^{-1} , \label{eq:MLE}
\end{equation}
where $n$ is the number of elements in a set $\{k_i \}$ whose
elements follow the power-law distribution $p(k) \sim k^{-\gamma}$,
and $k_\textrm{min}$ is the smallest element for which the power-law
behavior holds. Figure~\ref{MLE_deg} shows the change of the degree
exponent for the sampled networks from each network obtained by
numerical simulation for each method as we change the sampling
fraction $\alpha$.

For node sampling, we {\em fix} the number of sampled nodes and
select nodes randomly. In this case, the new degree distribution
$p'(k)$ of the sampled network is expressed as
\begin{equation}
p'(k) = \sum_{k_0 = k}^{n-1} p(k_0 ) {k_0 \choose k}{n - k_0 - 1
\choose n_s - k -1} \Big/ {n-1 \choose n_s - 1} ,
\label{new_degree_node_pj}
\end{equation}
where $p(k)$ is the degree distribution of the original network, $n$
is the number of nodes in the original network, and $n_s = \alpha n$
is the fixed number of sampled nodes. In the case that the number of
nodes in sampled networks is not fixed but only the probability
$\alpha$ with which individual nodes are selected is
given~\cite{Stumpf2005,Stumpf2005c}, Eq.~(\ref{new_degree_node_pj})
should be written as
\begin{equation}
p'(k) = \sum_{n_s = k+1}^n \sum_{k_0 = k}^{n-1} p(k_0 ) f(n_s ) {k_0
\choose k} \frac{\displaystyle {n - k_0 - 1 \choose n_s - k -1}}
{\displaystyle{n-1 \choose n_s - 1}} ,
\label{new_degree_node_pj_alpha}
\end{equation}
where the probability that $n_s$ number of nodes are chosen is
$f(n_s ) = {n \choose n_s} \alpha^{n_s} (1 - \alpha)^{n - n_s}$. If
the number of nodes is fixed, $f(n_s ) = \delta(n_s - \alpha n)$ and
Eq.~(\ref{new_degree_node_pj_alpha}) becomes
Eq.~(\ref{new_degree_node_pj}) with $n_s = \alpha n$. Even if the
number of nodes is not fixed, when the system size is large enough
to use the approximation $f(n_s ) \simeq \delta(n_s - \alpha n)$, we
can safely use Eq.~(\ref{new_degree_node_pj}).
Equation~(\ref{new_degree_node_pj}) can be further reduced by $n! /
(n-m)! \simeq n^m$ for $n \gg m$. Suppose $n, n_s \gg k_0 , k$. Then
\begin{equation}
\begin{array}{l}
\displaystyle{{n - k_0 - 1 \choose n_s - k -1} \Big/ {n-1 \choose
n_s - 1}}
\simeq n_s^k (n - n_s )^{k_0 - k}/ n^{k_0} \\ \\
= \displaystyle{\left(\frac{n_s}{n}\right)^k \left( 1 -
\frac{n_s}{n}\right)^{k_0 - k}} = \alpha^k (1 - \alpha)^{k_0 - k},
\end{array}
\end{equation}
which leads to the formula previously used in
Refs.~\cite{Cohen2000,Stumpf2005,Stumpf2005c}
\begin{equation}
p'(k) = \sum_{k_0 = k}^{\infty} p(k_0 ) {k_0 \choose k} \alpha^k
(1-\alpha)^{k_0 - k}. \label{new_degree_node}
\end{equation}

\begin{figure}
\includegraphics[width=0.4\textwidth]{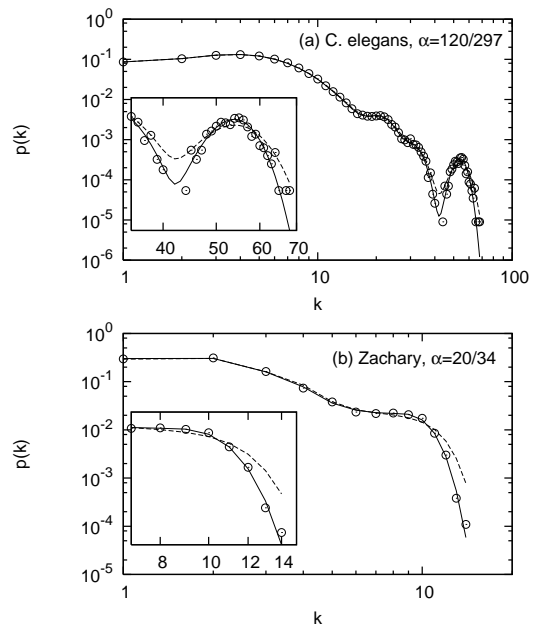}
\caption{Degree distribution for sampled networks of (a) {\em C.
elegans} neural network with $\alpha = 120/297$ and (b) Zachary
karate club network with $\alpha = 20/34$, obtained from the node
sampling. Empty circles are simulation results from $1000$ sampling
processes. Solid lines correspond to Eq.~(\ref{new_degree_node_pj})
and dashed lines to Eq.~(\ref{new_degree_node}). Insets show the
part of large degrees, where the difference between two formulae is
prominent, for each graph.} \label{SmallNet}
\end{figure}

The sizes of all the four networks studied in this paper are larger
than $5000$, and we have checked that
Eqs.~(\ref{new_degree_node_pj}) and (\ref{new_degree_node}) give
practically the same values of $p'(k)$ and are indistinguishable in
the graphs. For much smaller networks, on the other hand,
Eq.~(\ref{new_degree_node_pj}) actually predicts the degree
distribution of sampled networks better than
Eq.~(\ref{new_degree_node}). In Fig.~\ref{SmallNet}, we compare the
simulation results for two small networks, the nematode {\em C.
elegans} neural network~\cite{YYAhn2005} with $297$ nodes and $2359$
links and the Zachary karate club network~\cite{Girvan2002} with
$34$ nodes and $77$ links, with those two equations by substituting
the original degree distribution $p(k) = n_k / n$, where $n_k$ is
the number of nodes with degree $k$. The figure clearly shows that
Eq.~(\ref{new_degree_node_pj}) is more accurate.


The above equations turn out to be applied to the link sampling with
the same sampling fraction $\alpha$ as well. Here we can use the
technique in Ref.~\cite{Newman2005} to solve the bond percolation or
epidemic model. Suppose a node, which originally had $k_0$ links
before sampling, comes to have $k$ links. Because the random link
sampling chooses links uniformly, the probability of the node having
$k$ out of $k_0$ links is $p(k|k_0) = {k_0 \choose k} \alpha^k
(1-\alpha)^{k_0 - k}$. Consequently the probability that a node in
the sampled network has degree $k$ from all the possible original
degree $k_0$ is $p'(k) = \sum_{k_0 = k}^{\infty} p(k_0) p(k|k_0)$,
which leads us back to Eq.~(\ref{new_degree_node}). The fact that
those two sorts of sampling are described by the same equation is
also supported by Fig.~\ref{MLE_deg} showing the similar degree
exponent changes for both node and link sampling.

As Stumpf {\em et al.} point out in
Refs.~\cite{Stumpf2005,Stumpf2005c}, Eq.~(\ref{new_degree_node}) for
a power-law degree distribution $p(k) \sim k^{-\gamma}$ yields
deviation of $p'(k)$ from the original power-law form for quite
small sampling fraction $\alpha$. For moderate values of $\alpha$,
however, the deviation is not significant and we observe that the
tangent of $p'(k)$ in the log-log plot actually becomes steeper from
Eq.~(\ref{new_degree_node}), consistent with our numerical
observation about the node and link sampling as shown in
Fig.~\ref{MLE_deg}. To extract the degree exponents from
Eq.~(\ref{new_degree_node}), first we calculate the degree
distribution of the original networks by $p(k) = n_k / n$.
Substituting that $p(k)$ into Eq.~(\ref{new_degree_node}), we obtain
the degree distribution $p'(k)$ of the sampled networks
corresponding to a given sampling fraction $\alpha$. The degree
exponents from those $p'(k)$ in Fig.~\ref{MLE_deg} show good
agreement with the values from numerical simulation for both node
and link sampling cases.

\begin{figure}
\includegraphics[width=0.4\textwidth]{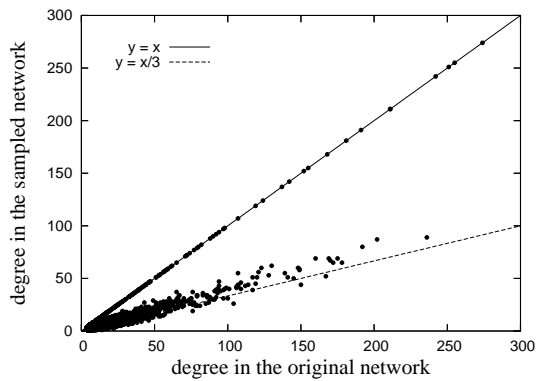}
\caption{Change of nodes' degree in BA network for snowball
sampling. The sampling fraction is $10000/30000$.}
\label{kk_snowball}
\end{figure}

On the contrary, it is found that a degree exponent decreases for
snowball sampling as we decrease the sampling fraction. By the
definition of snowball sampling, hubs are more likely to be selected
by this method. Furthermore, once a hub is picked, every node
connected to the hub is selected in the next step unless it belongs
to the previous layer. This characteristic of snowball sampling
tends to {\em conserve} the degrees of easily selected hubs, which
leads to the decrease of degree exponents by holding the ``tail'' of
the power-law degree distribution. Figure~\ref{kk_snowball} shows
the degrees in a sampled network obtained by snowball sampling, and
the nodes with large degree on the $y = x$ line clearly indicates a
tendency to choose hubs and conserve their degrees. Therefore, the
snowball sampling underestimates the degree exponent. In
Ref.~\cite{Clauset2004c}, they show that the traceroute sampling can
underestimate the degree exponent of a scale-free network by
undersampling the low-degree nodes relative to the high-degree ones.
In spite of the difference between the snowball and traceroute
sampling, both of these methods overrepresent hubs and have the same
``crawling'' character used to identify the nodes. We infer that the
decrease of degree exponents for both sampling methods is caused by
these similar features.

\begin{figure}
\includegraphics[width=0.45\textwidth]{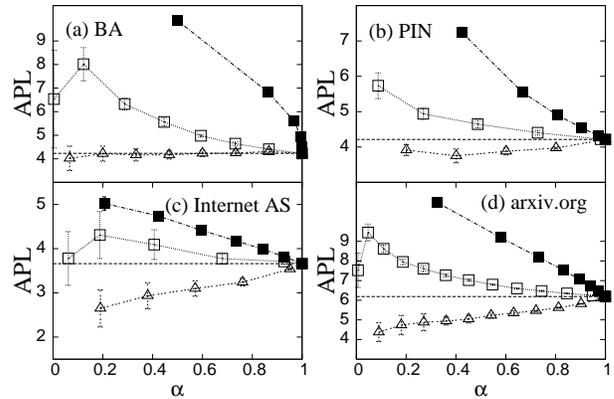}
\caption{Changes of APL for each network's sampled networks
according to the ratio $\xi$ of the size of giant component in the
sampled networks to that of the original ones, averaged over ten
independent realizations. Empty squares ($\square$) stand for node
sampling, filled squares ($\blacksquare$) for link sampling, and
empty triangles ($\vartriangle$) for snowball sampling. The
horizontal dashed lines are the values for the original APL of each
network, and the other lines are guides to the eyes.} \label{APL}
\end{figure}

We also check two closely related quantities, namely the average
degree and the average path length (APL) in the sampled networks.
APL is the average of shortest paths between all the pairs of nodes
in a network, often used as a measure of network efficiency. In
Fig.~\ref{APL}, we present APL of the {\em giant component} in the
sampled networks obtained by the numerical simulation. For snowball
sampling, APL decreases according to the decreased system size of
sampled networks. On the other hand, for node and link sampling, the
APL of a sampled network is larger than that of the original network
for not-so-small sampling fraction, even though the size of the
sampled network itself is smaller than the original one. As
presented earlier, for node sampling, the number of links is
proportional to the square of the number of the nodes, which leads
to $\langle k \rangle = 2l/n \propto n$, where $l$ and $n$ are the
numbers of links and nodes in a sampled network, respectively. This
suggests that the average degree in a sampled network decreases as
the sampling fraction becomes smaller. Obviously, for a given
network, APL decreases as the average degree increases. The
diminishment in the average degree, therefore, seems to have a
stronger effect on APL than the overall system size in this case.
Similar behavior of the average degree and APL is observed for link
sampling, but in this case it seems that the ``treelike'' structure
of sampled networks, related to the clustering coefficient discussed
later, is responsible for that behavior.

\subsection{Betweenness centrality distribution}

Betweenness centrality (BC or load), which measures the centrality
of a node by the traffic flow in a network, of node $b$ is defined
as
\begin{equation}
g_b = \sum_{i \neq j} \frac{C_b (i,j)}{C (i,j)},
\end{equation}
where $C (i,j)$ is the number of all the shortest pathways between a
pair of nodes $(i,j)$ and $C_b (i,j)$ is that of the shortest
pathways running through a node $b$~\cite{KIGoh2002}. It is known
that the BC distribution follows a power-law $p(g) \sim g^{-\eta}$
for scale-free networks~\cite{KIGoh2001,KIGoh2002}.

\begin{figure}
\includegraphics[width=0.45\textwidth]{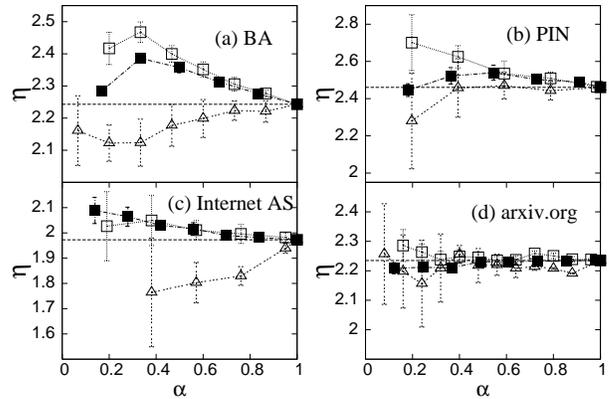}
\caption{Changes of BC exponent $\eta$ for each network's sampled
networks according to the sampling fraction $\alpha$, averaged over
ten realizations. Empty squares ($\square$) stand for node sampling,
filled squares ($\blacksquare$) for link sampling, and empty
triangles ($\vartriangle$) for snowball sampling. The horizontal
dashed lines are the values for the original exponent of each
network, and the other lines are guides to the eyes.}
\label{MLE_bc_BA}
\end{figure}

\begin{figure}
\includegraphics[width=0.4\textwidth]{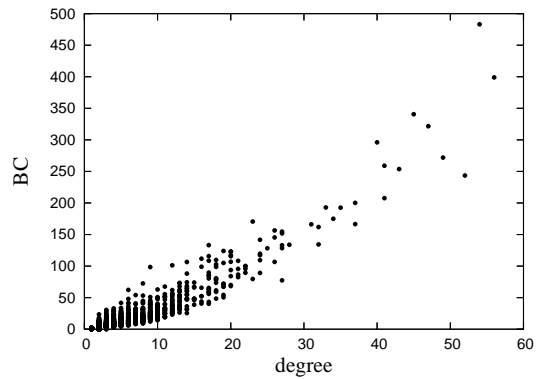}
\caption{Degree and BC of nodes in a sampled network of BA network
by node sampling. The sampling fraction is $10000/30000$. The value
of BC is rescaled by the number of nodes.} \label{kbc_node}
\end{figure}

Similar to the degree distribution, the BC distribution of sampled
networks also follows power-law well as do the original networks.
Figure~\ref{MLE_bc_BA} shows the change of the BC exponent, also
obtained by Eq.~(\ref{eq:MLE}), for each network and each sampling
method. Similar to the degree exponent case, in general, BC
exponents increase for node and link sampling and decrease for
snowball sampling as the sampling fraction gets lower.
Figure~\ref{MLE_bc_BA} bears a resemblance to Fig.~\ref{MLE_deg}
except for the case of arxiv.org, for which the BC exponent seems to
be conserved for all the sampling methods. The correlation between
degree and BC of nodes~\cite{Barthelemy2003b}, shown in
Fig.~\ref{kbc_node}, could explain the same direction of changes of
degree and BC exponents. For assortative networks such as arxiv.org
here, however, it is known that the degree-BC correlation is not
clear~\cite{KIGoh2003a}, which explains the different behavior in
Fig.~\ref{MLE_bc_BA}(d). Therefore, at least empirically, we expect
overestimation of a BC exponent by node and link sampling and
underestimation by snowball sampling.

\subsection{Assortativity}

The assortativity $r$, which measures the correlation between
degrees of node linked to each other, is defined as the Pearson
correlation coefficient of degrees between pairs of
nodes~\cite{Newman2002a}. Positive values of $r$ stand for the
positive degree-degree correlation which means that nodes with large
degrees tend to be connected to one another. Most social networks
have this positive degree correlation ({\em assortative} mixing),
including the arxiv.org network considered in this paper. On the
other hand, most biological and technological networks show negative
degree correlation $r < 0$ ({\em disassortative} mixing), including
PIN and Internet AS network here. If there is no degree correlation
among nodes (neutral), as in the case of BA model, the value of $r$
is in the vicinity of $0$.

\begin{figure}
\includegraphics[width=0.45\textwidth]{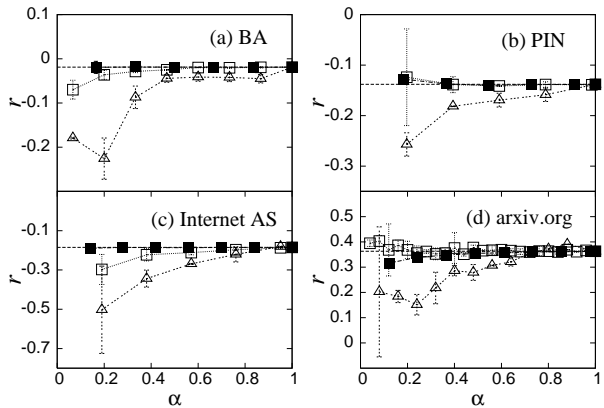}
\caption{Changes of assortativity $r$ for each network's sampled
networks according to the sampling fraction $\alpha$, averaged over
ten realizations. Empty squares ($\square$) stand for node sampling,
filled squares ($\blacksquare$) for link sampling, and empty
triangles ($\vartriangle$) for snowball sampling. The horizontal
dashed lines are the values for the original assortativity of each
network, and the other lines are guides to the eyes.}
\label{as_snowball}
\end{figure}

\begin{figure}
\includegraphics[width=0.45\textwidth]{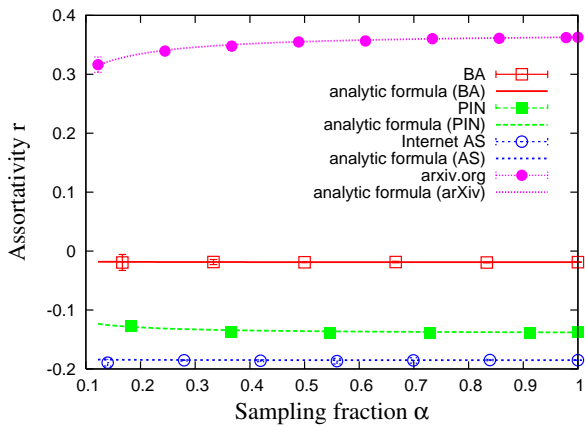}
\centering
\caption{Changes of assortativity $r$ under the link sampling
for our four datasets, and comparison with
Eq.~(\ref{link_assort_change}).}
\label{Sampling_figS}
\end{figure}

The change of assortativity for each network and each method is
shown in Fig.~\ref{as_snowball}. For node and link sampling, no
noticeable changes of assortativity in the sampled networks are
observed. Random choice of nodes or links appears to conserve
assortativity well for these two methods. Sampled networks from
snowball sampling, however, are shown to be more disassortative than
the original networks. This pattern is common no matter whether the
original network is assortative (arxiv.org), disassortative (PIN and
Internet AS), or neutral (BA). In Ref.~\cite{JDNoh2007a}, a formula
for the change of assortativity under the link sampling process
is presented as follows,
\begin{equation}
\label{link_assort_change}
r' = \frac{r}{\displaystyle{1 + \frac{1-\alpha}{\alpha} \left[
\frac{\langle k^2 \rangle / \langle k \rangle - 1}
{\langle k^3 \rangle / \langle k \rangle - ( \langle k^2 \rangle /
\langle k \rangle )^2 } \right]}},
\end{equation}
where $\langle k^n \rangle$ is the $n$th moment of the degree
of the original network.
Our data fit perfectly well with Eq.~(\ref{link_assort_change}), as
shown in Fig.~\ref{Sampling_figS}. In our datasets,
where the degree exponent $\gamma < 4$, $\langle k^3 \rangle$
dominates in Eq.~(\ref{link_assort_change}) and
$r' \simeq r$ in most cases, which is consistent with our
numerical data for the link sampling.

\begin{figure}
\includegraphics[width=0.4\textwidth]{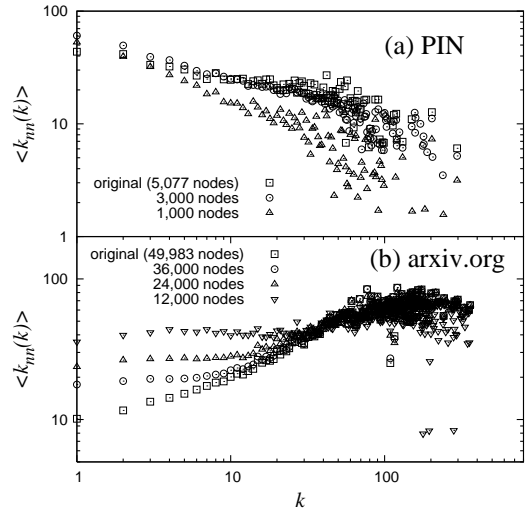}
\caption{$\langle k_{nn} (k) \rangle$ distribution for sampled
networks of (a) PIN, (b) arxiv.org by snowball sampling.}
\label{knn}
\end{figure}

There is another way to check the degree correlation, which is
measuring the quantity $\langle k_{nn} (k) \rangle = \sum_{k'} k'
p(k'|k)$, i.e., the average degree of nearest neighbors of nodes
with degree $k$~\cite{Satorras2001b}. Assortative mixing is
represented by a positive slope of the $\langle k_{nn} (k) \rangle$
graph, while the others by horizontal (neutral) or a negative slope
(disassortative). Figure~\ref{knn} shows the changes of these slopes
for $\langle k_{nn} (k) \rangle$ graphs of the sampled networks from
two kinds of original networks by snowball sampling. The slope
decreases, i.e., moves toward the negative value as the sampling
fraction gets lower for both disassortative Internet AS and
assortative arxiv.org.

We suggest that the more disassortative nature of sampled networks
compared with the original ones is due to the last layer of snowball
sampling method. In contrast to the conserved structure of the inner
layers, a considerable number of links are lost for the nodes in the
last layer. Meanwhile, hubs are likely to be selected for snowball
sampling. This separation of ``core'' and ``periphery'' part is seen
in Fig.~\ref{kk_snowball}, and the connections between hubs and
nodes of the last layer can reduce the value of assortativity. The
simulation shows that a sampled network containing the entire last
layer is more disassortative than the one where only parts of the
last layer are kept, which supports the hypothesis that the effect
of the last layer induces disassortative mixing. Therefore, we have
to be careful when measuring the assortativity for the network from
the snowball sampling.

\subsection{Clustering coefficient}

The clustering coefficient $C_i$ of node $i$ is the ratio of
the total number $y$ of the links connecting its nearest neighbors
to the total number of all possible links between all these
nearest neighbors~\cite{Dorogovtsev2001a},
\begin{equation}
C_i = \frac{2 y}{k_i (k_i - 1)} ,
\end{equation}
where $k_i$ is the degree of node $i$. The clustering coefficient of
a network is the average of this value over all the nodes $C =
\sum_i C_i / n$, where $n$ is the number of nodes. Most real
networks have much larger value of clustering coefficient than model
networks such as ER or BA network due to, e.g., the community or
modular structure.

\begin{figure}
\includegraphics[width=0.45\textwidth]{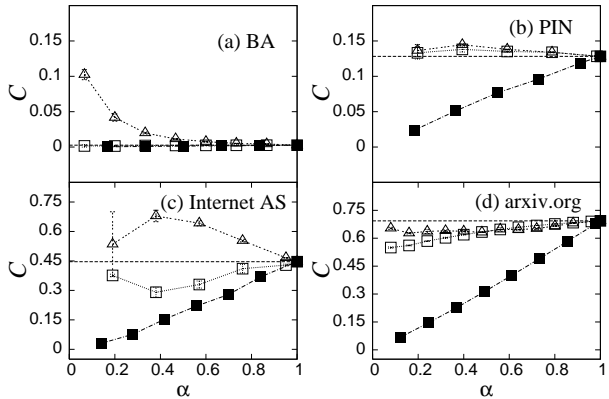}
\caption{Changes of clustering coefficient $C$ for each network's
sampled networks according to the sampling fraction $\alpha$,
averaged over ten realizations. Empty squares ($\square$) stand for
node sampling, filled squares ($\blacksquare$) for link sampling,
and empty triangles ($\vartriangle$) for snowball sampling. The
horizontal dashed lines are the values for the original clustering
coefficient of each network, and the other lines are guides to the
eyes.} \label{cc2_link}
\end{figure}

In Fig.~\ref{cc2_link}, we show the change of clustering coefficient
for each original network and each sampling method. For node and
snowball sampling, there is a little change of clustering
coefficient depending on networks. On the other hand, link sampling
prominently reduces the clustering coefficient. This effect is
obvious since the random omission of links, the reverse process of
link sampling, ``opens up triangles fast'' as stated in
Ref.~\cite{Kossinets2004}. The link sampling, therefore,
underestimates clustering coefficient of a network.

\section{discussion and conclusions}

\begin{table}
\begin{tabular}{ccccc}
\hline
\hline
 & Degree & BC &  & Clustering \\
 & Exponent & Exponent & Assortativity & Coefficient \\
 & $\gamma$ & $\eta$ & $r$ & $C$ \\
\hline
Node $\Downarrow$ & $\Uparrow$ & $\Uparrow$ & $=$ & $\Updownarrow$ \\
\hline
Link $\Downarrow$ & $\Uparrow$ & $\Uparrow$ & $=$ & $\Downarrow$ \\
\hline
Snowball $\Downarrow$ & $\Downarrow$ & $\Downarrow$ & $\Downarrow$ & $\Updownarrow$ \\
\hline
\hline
\end{tabular}
\caption{The changes of quantities in networks by each sampling
method. As the sampling fraction gets lower ($\Downarrow$ at the very right of
each sampling method indicates this),
$\Uparrow$ stands for increase, $\Downarrow$ for decrease, $=$ for the same,
and $\Updownarrow$ for depending on networks.}
\label{summary_table}
\end{table}

In this paper, we have studied the changes of well-known quantities
in complex networks for randomly sampled networks. Three kinds of
sampling methods are applied, and three representative real-world
networks, along with the BA model, are used as the original networks
for numerical investigation. We have measured four typical
quantities in sampled networks, which shows some characteristic
patterns in changes of the quantities for each sampling method.
Based on properties of sampling methods, possible explanations for
such changes as well as the mathematical analysis are provided. We
have also analyzed other networks than the scale-free ones such as
Erd\H{o}s-R{\'e}nyi random network~\cite{Erdos1959} and the growing
network without the preferential attachment~\cite{Barabasi1999}, and
the results show that the form of the degree distribution is
conserved for the node and link sampling in those cases, consistent
with the previous work~\cite{Stumpf2005c}.

Table~\ref{summary_table} summarizes the results. To check the
generality of the results, we also investigated the randomized
version of each network in a similar fashion. The randomized
networks were constructed by shuffling the links while conserving
only the degree distribution~\cite{Newman2002a}. We found the same
results with the original networks. The results in
Table~\ref{summary_table}, therefore, seem to hold for scale-free
networks in general and provide criteria for sampling method when
some specific quantity is supposed to be investigated by the
sampling. From another viewpoint, bias of some quantities can be
predicted if a specific sampling method used to identify a network
is known. If we are interested in the assortativity of a network,
for example, node or link sampling can give fairly accurate values.
For a clustering coefficient, on the other hand, the link sampling
method should be avoided.

Sampling problems should be taken into account for real network
research, but not much work has been done so far. Exploration of
other characteristics of complex networks or using other sampling
methods, rigorous analytic approaches, and establishing solid
principles by more systematic investigation could all be important
research topics for the future. We hope this work can make a
contribution to this direction of research.

\begin{acknowledgments}
We would like to thank Kwang-Il Goh for providing useful
information, and appreciate Yong-Yeol Ahn for helping us with the
link sampling formula. S.H.L. is grateful to Kim Bojeong Basic
Science Foundation and KAIST for generous help. This work was
supported by KOSEF through Grant No. R14-2002-059-01002-0 (P-J.K.)
and by R01-2005-000-1112-0 (H.J.).
\end{acknowledgments}


\begin{thebibliography}{99}
\bibitem{Traceroute} The traceroute command sends data packets
toward a certain destination on the Internet, and one collects the
traversed nodes and discovered paths during the process. Using such
processes repeatedly, the sampled networks of the whole Internet are
contructed at the AS or routers level.
\bibitem{pick_node} We verifed this relation $\beta \sim \alpha^2$ by
numerical simulation.
\end{thebibliography}
\end{document}